# DISCUSSION OF: TREELETS—AN ADAPTIVE MULTI-SCALE BASIS FOR SPARSE UNORDERED DATA

By Nicolai Meinshausen and Peter Bühlmann

*University of Oxford and ETH Zürich*

We congratulate Lee, Nadler and Wasserman (henceforth LNW) on a very interesting paper on new methodology and supporting theory. Treelets seem to tackle two important problems of modern data analysis at once. For datasets with many variables, treelets give powerful predictions even if variables are highly correlated and redundant. Maybe more importantly, interpretation of the results is intuitive. Useful insights about relevant groups of variables can be gained.

Our comments and questions include: (i) Could the success of treelets be replicated by a combination of hierarchical clustering and PCA? (ii) When choosing a suitable basis, treelets seem to be largely an unsupervised method. Could the results be even more interpretable and powerful if treelets would take into account some supervised response variable? (iii) Interpretability of the result hinges on the sparsity of the final basis. Do we expect that the selected groups of variables will always be sufficiently small to be amenable for interpretation?

**1. Treelets or hierarchical clustering combined with PCA.** A main part of the treelet algorithm achieves two main objectives:

(1) Variables are ordered in a hierarchical scheme. Highly correlated variables are typically "close" in the hierarchy.
(2) A basis on the tree is chosen. Each node of the tree is associated with a "sum" (and also a "difference" variable).

Clearly, treelets are more elegant than any method trying to achieve these two goals separately. As LNW write in Section 1: "The novelty and contribution of our approach is the simultaneous construction of a data-driven multi-scale orthogonal basis *and* a hierarchical cluster tree." We are left wondering, though, how different treelets are to the following scheme. First, variables are ordered in a hierarchical clustering scheme—for concreteness, under complete linkage and using similarities derived from absolute correlations as in (1). Second, a basis on the tree is found. For each node in the hierarchical clustering tree, the "sum" variable of the treelet algorithm would be replaced by the first PCA component of the variables represented by this node. Computationally, this scheme is clearly less efficient than the treelet algorithm, at least if implemented naively. Are there other benefits of taking steps (1) and (2) in one step as in the proposed treelet algorithm? It would be nice to see whether the tree structure of treelets differs substantially from a hierarchical cluster tree, and whether the treelets bases are very different from local PCA. Unfortunately, we did not obtain the treelet software from LNW, and that is the main reason why we did not pursue our own numerical experiments.

**2. Supervised and unsupervised basis selection.** In addition to contributions (1) and (2), treelets involve an additional step:

(3) Cut the hierarchical tree at some height, and work with the resulting basis. The chosen height is based on a clever score function; see formula (6).

The choice of the cut-point influences the "resolution" at which one is looking at the data. At one extreme (the leaves of the tree, "high resolution"), all variables are individual basis vectors. At the other extreme (the root of the tree, "low resolution"), basis vectors contain contributions from all variables, just like in global PCA. We understand the motivation behind the approach and the reported results seem to be very favorable. For supervised problems with a response, we are wondering if information in the response variable could be used more extensively to construct the treelet basis.

It is clear that a response variable should influence the choice of the basis. Take an example. If the signal-to-noise ratio (SNR) is very low, then one might be more inclined to work with "low resolution," as there is no hope of recovering the regression coefficients of individual variables. On the other hand, for high SNRs, it might very well be possible to single out individual variables as important. Information in the response variable could be used in various ways. Ranging from weak use of the response to stronger involvement:

(a) *Supervised choice of the cutoff height.* The cutoff of the tree can be influenced by the response. In fact, LNW used some supervised score function in Section 5.1 and also some cross-validation (and hence, supervised) approach in Section 5.3 to choose the best value for $K$, which in turn determines the cutoff value for the tree through criterion (6). Another possibility for finding the best cutoff in a supervised fashion would be to choose, instead of (6),

$$B_L = \underset{B_\ell : 0 \leq \ell \leq p-1}{\arg\min} \; CV(B_\ell),$$

where $CV(B_\ell)$ is the cross-validated loss of a favorite prediction method, using the orthogonal basis $B_\ell$ as predictor variables. Is it better to choose a value of $K$, and having then an associated best $K$-basis, or should we rather choose a best basis directly? Note that with the latter, we would also select features from the basis if the prediction method would do variable selection, for example, the Lasso or tree-based methods including boosting or random forests.

(b) *Nonuniform cutoff height.* For a given tree, it is not obvious why cutting at a single height is necessarily optimal. As an example, take 2 predictor variables $x_i$ and $x_j$ with $i \neq j$ who are quite correlated and both of them are strongly relevant for prediction. They will tend to be merged quite early in the tree, but we would like to keep them separate for interpretation and best predictive performance (while we would like to merge as early as possible less correlated clusters of variables that only have a weak influence on the response).

Instead of cutting the tree at a single height, it might be more advantageous to start toward the root node of the tree. If a given cluster of variables turns out to be important, one could try to add—in a forward selection manner—basis elements from its sub-clusters. If descending deeper into the tree at a particular node improves prediction considerably, one would keep descending and stop otherwise. The selected tree height would not be uniformly the same. The resolution would be high in directions of strong signal and low in directions of weak signal. For related procedures, see also Meinshausen (2008) or Goeman and Mansmann (2008). And also "supervised harvesting" [Hastie, Tibshirani, Botstein and Brown (2001)] has the property that features at different levels of a hierarchical cluster tree are selected.

(c) *Supervised tree growth.* Take again the example in (b) of two rather correlated predictor variables, who are merged quite early in the tree but contribute both strongly to the response. A more principled way of dealing with the issue would be to make the *construction* of treelets, that is, the tree and the bases, supervised. Is it possible? [Besides doing the obvious, viz., to include the response $y$ as another variable, i.e., considering new data $\tilde{x} = (y, x)$.] To our knowledge, there are not many methods for "supervised grouping." It seems to us that among the references in LNW, only the method in Dettling and Bühlmann (2004) remains as "truly supervised," while the elastic net approach in Zou and Hastie (2005), which is supervised, is not extracting a group structure.

We think that it would be worthwhile to extend treelets in the direction of a truly supervised algorithm both for improved prediction performance and better interpretability.

**3. Interpretability.** One attractive property of treelets is the sparsity of the solution (sparsity is here to be understood as few variables entering a basis vector). Compared with global PCA, which includes contributions from all variables into every basis vector, treelet basis vectors contain in general only a few variables in each basis vector. This increases the interpretability of results dramatically.

There is clearly a tradeoff, though: increasing the sparsity increases interpretability by performing variable selection among the treelet features. Increasing sparsity increases at the same time, however, the variance of the solution. Making the results very sparse carries, in general, the risk that the results are unstable. We might see a completely different result on repeated measurements (or on repeated bootstrap samples). We would thus like to make the results "as sparse as possible, but not any sparser." A very sparse yet unstable result is not suitable for interpretation either.

What should we do if the selected groups of variables will be too large for interpretation? For example, groups of genes of size more than 20 are often an idea attractive to statisticians or computer scientists, but it is very likely that such large groups will never be validated by biological experiments. Is the solution as simple as cutting the tree at a level such that the group size is bounded by a value which is desired for a specific application?

Bounding the maximal group size can potentially render the algorithm unstable. As a possible solution to the sparsity–stability tradeoff, we can cut the tree at a height that gives maximal sparsity of results under the condition that the obtained groups of variables are—in some sense—stable under permutations of the data. LNW show in Figure 3 some bootstrap confidence bands which are supported by some asymptotic theory in Section 3.1. It would be interesting to have a more complete way of visualizing the stability of the treelet procedure.

**4. Conclusions.** We think that treelets are a very interesting and promising proposal for high-dimensional modern data analysis. Open-source high quality software would be desirable: it would help promoting the method to a large community of users and researchers and it would allow reproducibility of results.

Department of Statistics  
University of Oxford  
1 South Parks Road  
Oxford OX1 3TG  
United Kingdom  
E-mail: meinshausen@stats.ox.ac.uk

Seminar für Statistik  
ETH Zürich  
Leonhardstrasse 27  
CH-8092 Zürich  
Switzerland  
E-mail: buhlmann@stat.math.ethz.ch